# Simultaneous generation of Raman-assisted Soliton Microcombs and Tunable Multi-chromatic Raman Microlasers in Single Monolithic Thin-film Lithium Niobate Microrings

*Yingnuo Qiu,[1,4] Renhong Gao,[2,§] Chuntao Li,[2,3] Yixuan Yang,[1,4] Xinzhi Zheng,[2,3] Guanghui Zhao,[1] Xiaochao Luo,[1,4] Qifeng Hou,[1] Lingling Qiao,[1] Min Wang,[2,3] Jintian Lin,[1,4,*] and Ya Cheng[2,3]*

[1]State Key Laboratory of High Field Laser Physics and CAS Center for Excellence in Ultra-Intense Laser Science, Shanghai Institute of Optics and Fine Mechanics (SIOM), Chinese Academy of Sciences (CAS), Shanghai 201800, China

[2]The Extreme Optoelectromechanics Laboratory (XXL), School of Physics and Electronic Science, East China Normal University, Shanghai 200241, China

[3]State Key Laboratory of Precision Spectroscopy, East China Normal University, Shanghai 200062, China

[4]Center of Materials Science and Optoelectronics Engineering, University of Chinese Academy of Sciences, Beijing 100049, China

[§]rhgao@phy.ecnu.edu.cn

*jintianlin@siom.ac.cn

High-performance integrated broadband coherent light sources are essential for advanced applications in high-bandwidth data processing and chip-scale metrology, yet remain challenging. In this study, we demonstrate a monolithic Z-cut lithium niobate on insulator (LNOI) microring platform that enables simultaneous generation of tunable multi-chromatic microlasers and Raman-assisted soliton microcombs. Exploiting the strong Raman activity and high second-order nonlinearity of LNOI, we engineered a dispersion-optimized microring with a loaded Q factor of $3.86\times10^6$, facilitating on-chip efficient broadband coherent light source. A novel phase-matching configuration with all the waves of the same ordinary polarization was realized for the first time in this platform, feasibly enabling modal-phase matched Raman-quadratic nonlinear processes that extend lasing signals into the visible spectrum. Under continuous-wave laser pumping at 3.73 mW in the telecom band, we achieved a Raman-assisted soliton comb centered at 1624.49 nm with record-low pump threshold on the LNOI platform. Concurrently, multi-chromatic Raman lasing outputs were observed at ~1700, ~813, and ~535 nm within the same microring. The system exhibited efficient wavelength tuning of these multi-chromatic laser signals through a 5 nm shift in pump wavelength. This work represents a significant advance in integrated photonics for versatile optical signal generation.

**1. Introduction**

With the remarkable ability to harness the outstanding optical, electronic, and acousto-optic properties of lithium niobate on insulator (LNOI) in compact micro/nano-structures, a variety of chip-scale photonic devices have been developed for applications ranging from high-bandwidth information processing, miniatured precision metrology, to quantum technology.[1-21] The substantial progresses in LNOI photonics have spurred interest in approaches to develop both tunable microlasers with multi-chromatic outputs and soliton microcombs to extend coherent spectral bandwidth for boosting on-chip functionalities.[8-12,14,15,17,22] Generally, the soliton microcombs are generated via Kerr nonlinearity in anomalous-dispersion LNOI microresonators,[9,12,23-25] whilst the tunable multi-chromatic laser sources involve using the nonlinear optical properties of lithium niobate (LN) like stimulated Raman scattering (SRS) and high second-order nonlinearity ($\chi^{(2)}$) for Raman lasing and quadratic frequency conversion.[10,26,27] Therefore, the tunable multi-chromatic microlasers and soliton microcombs have been demonstrated separately on photonic devices so far. Moreover, because of the negative uniaxial property of LN crystal, phase matching configuration with all the waves of the same ordinary (o) polarization which is inaccessible in bulk LN in a spectral region of normal dispersion (<1900 nm),[28] has yet to be demonstrated for providing new freedom for quadratic nonlinear processes.

In this work, a dispersion engineered LNOI microring was exploited to simultaneously generate the Raman-assisted soliton microcombs and multi-chromatic Raman lasers, which is enabled by the collaborative excitation of two strong Raman vibration phonon branches and high $\chi^{(2)}$ for Raman-quadratic nonlinear frequency conversion. A novel phase matching configuration with all the waves of the same ordinary (o) polarization was proposed and realized for the first time on the LNOI platform, so as to trigger Raman-quadratic process for generation of multi-chromatic Raman lasing sources. Under continuous-wave pumping in the telecom band with a pump level exceeding 3.73 mW, a Raman-assisted soliton comb centered at 1624.486 nm was observed. Furthermore, multi-wavelength Raman-laser outputs at approximately 1700 nm, 813 nm, and 535 nm were generated within the same microring, significantly extending the integrated coherent spectral bandwidth. Finally, by leveraging the flexible spectral tunability of stimulated Raman scattering, these multi-chromatic Raman signals achieved a tuning bandwidth of several nanometers by varying the pump wavelength of a 5 nm range.

## 2. Raman nonlinearity in a monolithic thin-film lithium niobate (TFLN) microring

### 2.1 Fabrication of the monolithic microrings on lithium niobate on insulator (LNOI) platform

The monolithic microrings for this experiment were fabricated using a commercial Z-cut LNOI wafer (Jinan Jingzheng Electronics Technology Co., Ltd.). The wafer consists of a lithium niobate (LN) thin film (700-nm thick), a silicon dioxide ($SiO_2$) buffer layer (4.7-μm thick), and a silicon handle (500-μm thick). Microring resonators supporting a few high-Q spatial whispering gallery modes (WGMs) were successfully fabricated through femtosecond laser lithography-assisted chemo-mechanical etching (PLACE) technique.[29] The specific process flow consists of 4 steps as follows: First, a 600-nm thick chromium (Cr) layer was deposited on the surface of the LNOI wafer. Subsequently, the femtosecond laser was utilized to selectively remove the Cr layer with a spatial resolution of 200 nm, so as to form the microring-pattern hard mask. The patterns were then transferred into the LN thin film via chemo-mechanical polishing (CMP) with an etching depth of 500 nm of the exposed LN thin film. Finally, the Cr mask was removed by wet chemical etching, followed by a secondary CMP step to eliminate surface defects and optimize structural flatness. The LNOI microring resonator was fabricated with a radius of 200 μm, and a thickness of 640 nm, which is side-coupled with a ridge waveguide. The inset of Fig. 2(a) shows a false-colored scanning electron microscope (SEM) image of the microring, demonstrating good structural quality and surface morphology. The coupling gap between the bus waveguide and the microring was controlled to approximately 5.5 μm to make a tradeoff between the coupling efficiency and coupling loss.

## 2.2 Experimental setup for cavity-mode excitation, characterization, and formation of on-chip broadband Raman-nonlinearity coherent light sources

As shown in Fig. 1, to excite the cavity modes in the fabricated TFLN microring resonator, we employed a narrow-linewidth tunable laser (short-term linewidth < 200 kHz, Model: TLB-6728, New Focus Inc.) as the pump source. Its output power was regulated by an in-line variable optical attenuator (VOA). The pump light was edge-coupled into the device under test (i.e., photonic chip composed of microrings side-coupled with waveguides) via a lensed fiber, whose position was precisely controlled by a 3D piezoelectric stage with a resolution of 20 nm. An optical microscope imaging system was set up above the photonic chip to real-time monitor the end-butt coupling between the lensed fiber and the waveguide, comprising an objective lens with a numerical aperture (NA) of 0.28, optional optical filters, and an infrared (IR)/visible charge-coupled device (CCD) camera. The output signal was collected by another lensed fiber of the same model for both nonlinear spectrum analysis and mode structure characterization.

For mode structure characterization, the power level of the pump laser injected into the microring was adjusted as low as 5 µW to avoid exciting any nonlinear process and thermo-optic effect, and the output signal was directly sent to a photodetector (Model 1811, New Focus Inc.). The photodetector (PD) was then connected to an oscilloscope (Model: MOS 64, Tektronix Inc.) to monitor the coupling of resonant modes within the TFLN microring for characterizing the mode structure and the Q factors of the modes. For nonlinear spectrum analysis, the pump laser was regulated to a certain wavelength with above-threshold power to excite the Raman nonlinearity in the microring, and the output nonlinear signals coupled out of the microring via the lensed fiber were directed into an optical spectrum analyzer (Models: AQ6370D/AQ6375B, Yokogawa Electric

Corporation) and a spectrometer (Model: USB2000, Ocean Optics, Inc.) to detect optical emission spanning from 330 nm to 2300 nm. Moreover, the nonlinear optical signals emitted from the surface of the microring were captured by the real-time imaging system to record the spatial intensity profile of the excited modes. The pump power used in the calculations of the SRS threshold refers to the on-chip incident power, accounting for the fiber-to-chip coupling efficiency. Correspondingly, the output powers of the generated Raman-nonlinearity lasing signals used for calculating the conversion efficiency have already been corrected to account for the coupling loss of the lensed fiber.

**2.3 Simultaneous generation of Raman-assisted soliton frequency combs and tunable multi-color laser output in the monolithic TFLN microring**

**2.3.1 Mode structures in the Microring and integrated dispersion**

Figure 2(b) displays the typical mode structure in the telecom band, where the free spectral range (FSR) of the fundamental transverse electric (TE$_0$) WGMs was ~0.81 nm. The TE$_0$ mode at ~1559.14 nm was ascribed to TE$_{0,1626}$, with the underscripts ($n$, $m$) of the mode denoting the radial mode number and azimuthal mode number, respectively. This mode possesses ultra-high Q factor, which was measured as high as $3.86 \times 10^6$, as depicted in Fig. 2(c). The simulated integrated dispersion ($D_{int}/2\pi$) curves of the fundamental transverse electric (TE$_0$) WGMs around the wavelength of 1624 nm are plotted in Fig. 2(d), exhibiting anomalous dispersion. As a result, the Raman-assisted four-wave mixing is allowed to be excited through triggering the SRS process around 1624 nm.

**2.3.2 Multi-chromatic Raman laser outputs in the microring**

When the input wavelength $\lambda_p$ was adjusted to pump the TE$_{0,1626}$ mode at ~1559.14 nm, forward-propagating Stokes SRS signals were detected at wavelengths around 1624 nm and 1700 nm, as depicted in Fig. 3(a). Referring to the Raman spectrum of lithium niobate (LN) crystal, we identified the spectral line at the wavelength $\lambda_R$ of 1699.55 nm as a first-order Raman signal near the pump wavelength associated with the (8E TO) Raman-active phonon branch,[30] corresponding to a Raman shift of 580 cm$^{-1}$. And this SRS signal was resonant with the TE$_{0,1470}$ mode, whose electric field profile is plotted in the inset of Fig. 3(b). The side-mode suppression ratio (SSR) of this SRS signal reaches 48 dB when the pump power was 2.93 mW, as shown in Fig. 3(b). Figure 3(c) plots the Raman laser output power varied with increasing pump power in the small-signal gain regime, showing a linear dependence with a conversion efficiency of 27.75% and a threshold of 2.56 mW.

And the Stokes SRS signal observed at 1624.49 nm corresponds to a Raman shift of 265 cm$^{-1}$, which was resonant with the TE$_{0,1550}$ mode identified by theoretial simulation. This is a first-order Raman signal near the pump wavelength associated with the (3E TO) Raman-active phonon branch.[30] The variation of the output power of this signal with increasing pump power in the small-signal gain regime is shown in Fig. 3(d), exhibiting a linear growth. The conversion efficiency and threshold were determined to 42.48% and 2.87 mW, respectively. When the pump power was raised to 3.11 mW, Raman-assisted four-wave mixing, which is enabled by the anomalous dispersion in this waveband, appeared in the spectrum, as shown in Fig. 3(e). When the pump power was further raised to 3.73 mW, spectral broadening was observed, showing a frequency comb spanning a spectral range from 1617 nm to 1629 nm, as presented in Fig. 3(f). Moreover, the generated comb envelope can be well-fitted by a sech² function, confirming that this constitutes a stable soliton frequency comb generated via Raman-assisted four-wave mixing. This pump threshold of 3.73 mW for soliton microcomb generation is a three-fold improvement compared to the state of the art on the same material platform.[12,24]

At this pump level, obvious visible-light emission from the top surface of the microring was also observed with naked eyes, which was recorded by the real-time optical microscope imaging system, as shown in Fig. 4(a). A near-infrared nonlinear signal (NIR-R) was detected at 813.16 nm. The output power of this NIR-R signal was measured to be 50.36 μW, showing a high SSR of 30 dB, as shown in Fig. 4(b).

The output power of this NIR-R signal increased linearly with the square of the aforementioned pump power, exhibiting a conversion efficiency of 1.89%/mW and a threshold of 3.37 mW, as shown in Fig. 4(c). These characteristics align well with the nature of the stimulated two-photon hyper-Raman scattering process.[26] Using an 800 nm long-pass filter and a 950 nm short-pass filter to filter other shortwave signals, the spatial mode intensity distribution of the NIR-R mode was captured separately by the visible CCD camera, as shown in the inset of Fig. 4(c).

In addition to this NIR-R signal, another visible signal (V-R) was observed in the spectrum of Fig. 4(d), located at 535.02 nm. Considering that the wavelength $\lambda_{R''}$ of this V-R signal is equal to $\frac{1}{1/\lambda_p + 1/\lambda_{R'}}$, this V-R signal was generated through the sum-frequency generation (SFG) of the pump light and the NIR-R signal. There is no other signal in 450-600 nm waveband. The optical microscope imaging system was used to record the intensity profile of this signal emitted from the surface of the macroring, by filtering other signals with a 500 nm long-pass filter and a 600 nm short-pass filter, exhibiting green light emission, as shown in the upper inset of Fig. 4(d). Meanwhile, after filtering other signals of the wavelength longer than 450 nm via a 450 nm short-pass filter, a bright blue visible light emission from the microring was captured, as shown in the lower inset of Fig. 4(f). This signal is likely a new Raman signal generated via a three-photon hyper-Raman process.[31] However, due to severe signal attenuation through the optical fiber, the signal is too weak to be detected by the optical spectrum meter.

**2.3.3 Analysis of phase matching condition**

It is necessary to unveil the modal phase matching configuration in Raman-quadratic nonlinear process, where phase matching conditions are crucial for optical parametric nonlinear interactions. [7,13,16,19-21,31-36] In the Z-cut LNOI microring, the transverse-magnetically (TE) polarized waves correspond to ordinary (o) waves. In consideration of the negative uniaxial property and normal material dispersion of LN crystal, it is impossible to achieve the phase matching configuration with all the waves of the same ordinary polarization for quadratic nonlinear process such as SFG termed "o + o → o" within the material normal-dispersion spectral range (i.e., wavelength <1900 nm). However, in contrast to bulk LN crystal, dispersion of LNOI microring can be changed by trimming the geometrical structure of the microring. In our microring, the wavevector phase matching for the formation of the NIR-R Raman lasing signal at 813.16 nm was fulfilled by dispersion engineering, which can be expressed as

$$m_{R\prime} = m_P + m_R, \qquad (1)$$

where $m_{R\prime}$, $m_P$, and $m_R$ are the azimuthal mode numbers of the NIR-R signal, pump light, and the SRS signal at 1699.55 nm, respectively. And the effective second-order nonlinear coefficient for this Raman-quadratic process can be expressed as

$$d_{eff} = 3d_{22} \cdot sin^2\phi \cdot \cos\phi - d_{22} \cdot cos^3\phi, \qquad (2)$$

where $\phi$ is the angle between the wave vector $\vec{k}$ and the horizontal coordinate axis $x$ which is schemtially defined in the inset of Fig. 1, and $d_{22}$ is the nonlinear susceptibility of LN. Consequently, the NIR-R signal at 813.16 nm was generated by the novel model phase matched SFG of the pump light and the SRS signal at 1699.55 nm with the waves of the same ordinary polarzation, by leveraging the second-order nonlinear susceptibility of $d_{22}$ of LN to feasibly providing new degree freedom to boost phase-matched Raman-quadratic process.

**2.3.4 Tunable multi-chromatic laser signals**

To investigate the tunability of the Raman laser, we continuously tuned the pump wavelength at an interval of ~0.81 nm (to excite different WGMs of the same pump mode family of $TE_0$), while maintaining the pump power at 3.73 mW, which resulted in drift of both the SRS and the NIR-R signals. Figures 5(a) and (c) show that when the pump wavelength was tuned from 1555.69 nm to 1560.51 nm, the stimulated Raman laser exhibited a tuning range from 1695.72 nm to 1700.56 nm, and the NIR-R laser exhibited a tuning range from 811.33 nm to 813.75 nm. And both the wavelengths of the stimulated Raman laser and the NIR-R laser linearly red-shifted with the increasing pump wavelength, as confirmed in Figs. 5(b) and (d).

**3. Conclusion**

In conclusion, concurrent generation of Raman-assisted soliton microcombs and multi-color Raman lasers was demonstrated in a dispersion-engineered LNOI monolithic microring resonator. This broadband multi-chromatic coherent light source was generated by the synergistic excitation of two prominent Raman-active phonon branches alongside the strong $\chi^{(2)}$ inherent to LN enabled by the novel modal phase matching configuration with all the ordinary-polarization waves. This work paves the way for the realization of fully integrated tunable multi-chromatic coherent light sources, which will further promote the development of applications in precision spectroscopy and sensing.

**References**
[1] J. Lin, F. Bo, Y. Cheng, J. Xu, *Photon. Res.* **2020**, 8, 1910.
[2] Z. Xie, F. Bo, J. Lin, H. Hu, X. Cai, X.-H. Tian, Z. Fang, J. Chen, M. Wang, F. Chen, Y. Cheng, J. Xu, S. Zhu, *Adv. Phys. X* **2024**, 9, 2322739.
[3] Y. Jia, L. Wang, F. Chen, *Appl. Phys. Rev.* **2021**, 8, 011307.
[4] J. J. Chakkoria, A. Dubey, A. Mitchell, A. Boes, *Opto-Electronic Adv.* **2025**, 8, 240139.
[5] R. Gao, N. Yao, J. Guan, L. Deng, J. Lin, M. Wang, L. Qiao, W. Fang, Y. Cheng, *Chin. Opt. Lett.* **2022**, 20, 011902.
[6] C. Wang, M. Zhang, M. Yu, R. Zhu, H. Hu, M. Loncar, *Nat. Commun.* **2019**, 10, 978.
[7] J. Lin, N. Yao, Z. Hao, J. Zhang, W. Mao, M. Wang, W. Chu, R. Wu, Z. Fang, L. Qiao, W. Fang, F. Bo, Y. Cheng, *Phys. Rev. Lett.* **2019**, 122, 173903.
[8] J. Han, M. Li, R. Wu, J. Yu, L. Gao, Z. Fang, M. Wang, Y. Liang, H. Zhang, Y. Cheng, *Opto-Electronic Sci.* **2025**, 4, 250004.
[9] S. Wan, P.-Y. Wang, M. Li, R. Ma, R. Niu, F.-W. Sun, F. Bo, G.-C. Guo, C.-H. Dong, *Nat. Commun.* **2015**, 16, 4829.
[10] Y. Zhao, X. Liu, K. Yvind, X. Cai, M. Pu, *Commun. Phys.* **2023**, 6, 350.
[11] M. Yu, Y. Okawachi, R. Cheng, C. Wang, M. Zhang, A. L. Gaeta, M. Lončar, *Light Sci. Appl.* **2020**, 9, 9.
[12] B. Fu, R. Gao, N. Yao, H. Zhang, C. Li, J. Lin, M. Wang, L. Qiao, Y. Cheng, *Opto-Electronic Adv.* **2024**, 7, 240061.
[13] Y. Zhang, H. Li, T. Ding, Y. Huang, L. Liang, X. Sun, Y. Tang, J. Wang, S. Liu, Y. Zheng, X. Chen, *Optica* **2023**, 10, 688.
[14] J. Lin, S. Farajollahi, Z. Fang, N. Yao, R. Gao, J. Guan, L. Deng, T. Lu, M. Wang, H. Zhang, W. Fang, L. Qiao, Y. Cheng, *Adv. Photon.* **2022**, 4, 036001.
[15] J. Ling, Z. Gao, S. Xue, Q. Hu, M. Li, K. Zhang, U. A. Javid, R. Lopez-Rios, J. Staffa, Q. Lin, *Nat. Commun.* **2024**, 15, 4192.
[16] Z.-Y. Wang, X. Wu, X. Xiong, C. Yang, Z. Hao, Q.-F. Yang, Y. Hu, F. Bo, Q.-T. Cao, Y.-F. Xiao, *Sci. Adv.* **2025**, 11, eadu7605.
[17] Z. Li, R. N. Wang, G. Lihachev, J. Zhang, Z. Tan, M. Churaev, N. Kuznetsov, A. Siddharth, M. J. Bereyhi, J. Riemensberger, T. J. Kippenberg, *Nat. Commun.* **2023**, 14, 4856.
[18] Z. Jiang, C. Fang, X. Ran, Y. Gao, R. Wang, J. Wang, D. Yao, X. Gan, Y. Liu, Y. Hao, G. Han, *Opto-Electronic Adv.* **2025**, 8, 240114.

[19] J. Lu, M. Li, C.-L. Zou, A. A. Sayem, H. X. Tang, *Optica* **2020**, 7, 1654.

[20] H.-Y. Liu, M. Shang, X. Liu, Y. Wei, M. Mi, L. Zhang, Y.-X. Gong, Z. Xie, S. N. Zhu, *Adv. Photon. Nexus* **2022**, 2, 016003.

[21] J. Hou, J. Zhu, R. Ma, B. Xue, Y. Zhu, J. Lin, X. Jiang, X. Chen, Y. Cheng, L. Ge, Y. Zheng, W. Wan, *Laser Photonics Rev.* **2024**, 18, 2301351.

[22] Q. Luo, F Bo, Y. Kong, G. Zhang, J. Xu, *Adv. Photon.* **2023**, 5, 034002.

[23] C. Yang, S. Yang, F. Du, X. Zeng, B. Wang, Z. Yang, Q. Luo, R. Ma, R. Zhang, D. Jia, Z. Hao, Y. Li, Q. Yang, X. Yi, F. Bo, Y. Kong, G. Zhang, J. Xu, *Laser Photonics Rev.* **2023**, 17, 2200510.

[24] Y. He, Q.-F. Yang, J. Ling, R. Luo, H. Liang, M. Li, B. Shen, H. Wang, K. Vahala, Q. Lin, *Optica* **2019**, 6, 1138.

[25] T. J. Kippenberg, A. L. Gaeta, M. Lipson, M. L. Gorodetsky, *Science* **2018**, 361, eaan8083.

[26] G. Zhao, J. Lin, B. Fu, R. Gao, C. Li, N. Yao, J. Guan, M. Li, M. Wang, L. Qiao, Y. Cheng, *Laser Photonics Rev.* **2024**, 18, 2301328.

[27] Y. He, X. Yan, J. Wu, X. Liu, Y. Chen, X. Chen, *Opt. Lett.* **2024**, 49, 4863.

[28] R. W. Boyd, *Nonlinear Optics (3rd Edition, Academic Press, 2009)*, Chapter 2.

[29] C. Li, J. Guan, J. Lin, R. Gao, M. Wang, L. Qiao, L. Deng, Y. Cheng, *Opt Express* **2023**, 31, 31556.

[30] V. S. Gorelik, S. D. Abdurakhmonov, N. V. Sidorov, M. N. Palatnikov, *Inorganic Materials* **2019**, 55, 524.

[31] C. Li, N. Yao, H. Yu, J. Lin, R. Gao, J. Deng, J. Guan, L. Qiao, Y. Cheng, *Phys. Rev. Lett.* **2025**, 134, 213801.

[32] X. Luo, C. Li, X. Huang, J. Lin, R. Gao, Y. Yao, Y. Qiu, Y. Yang, L. Wang, H. Yu, Y. Cheng, Nat. Commun. 2025 (DOI: 10.1038/s41467-025-66647-2).

[33] I. Breunig, *Laser Photonics Rev.* **2016**, 10, 569.

[34] T. Yuan, J. Wu, Y. Liu, X. Yan, H. Jiang, H. Li, Z. Liang, Q. Lin, Y. Chen, X. Chen, *Sci. China-Phys. Mech. Astron.* **66**, 284211 (2023).

[35] Z. Xie, X. Lv, and S. Zhu, Sub-coherence-length nonlinear optical manipulation via twist phase matching. *Science Bulletin* **69**, 1170 (2024).

[36] T. Yuan, J. Wu, X. Wang, C. Chen, H. Li, B. Wang, Y. Chen, X. Chen, *Adv. Photon.* **6**, 056012 (2024).

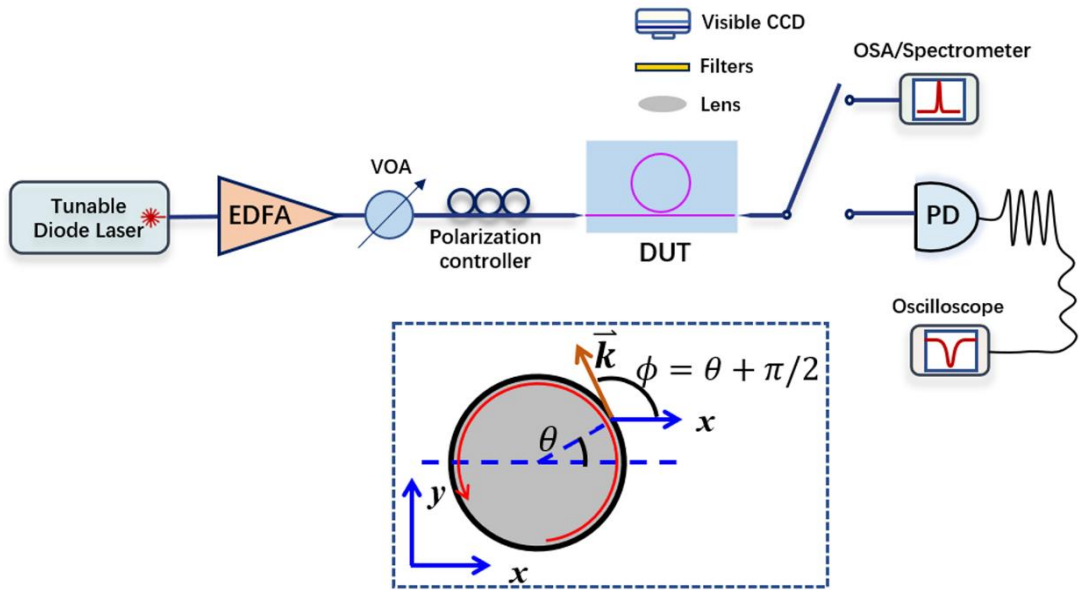

**Figure 1.** The experimental setup for the formation of integrated tunable broadband multi-chromatic coherent light sources in the LNOI microring based on Raman-nonlinearity. Inset: waves circulating in the microring with varying azimuthal angle $\theta$. DUT: device under test; $\vec{k}$: the wave vector of the wave; $\phi$: the angle between the wave vector and the horizontal axis in the microring plane.

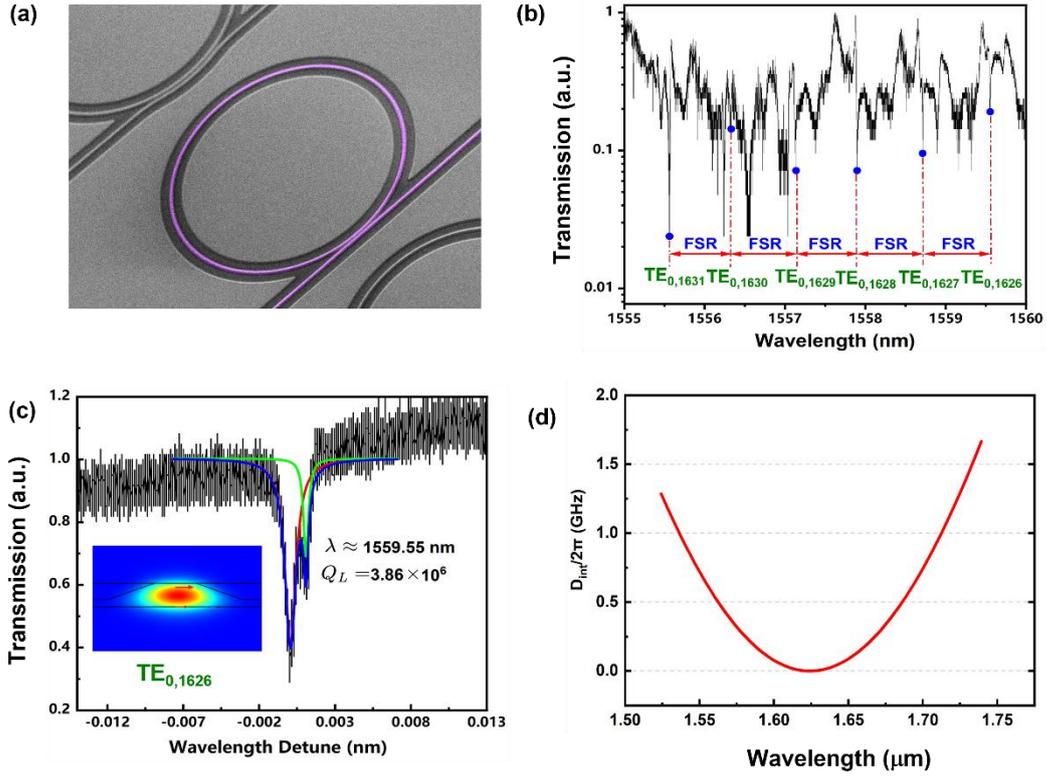

**Figure 2.** (a) False-colored scanning electron microscope (SEM) image of the monolithic microring. (b) Mode structure of the microring in the telecom band, where the free spectral range (FSR) of the TE$_0$ familiy was ~0.81 nm. (c) Loaded Q factor of the pump mode. Inset: Simulated electric field profile of the pump mode TE$_{0,1626}$. (d) Simulated integrated dispersion ($D_{int}/2\pi$) curve of the TE$_0$ modes around 1624 nm.

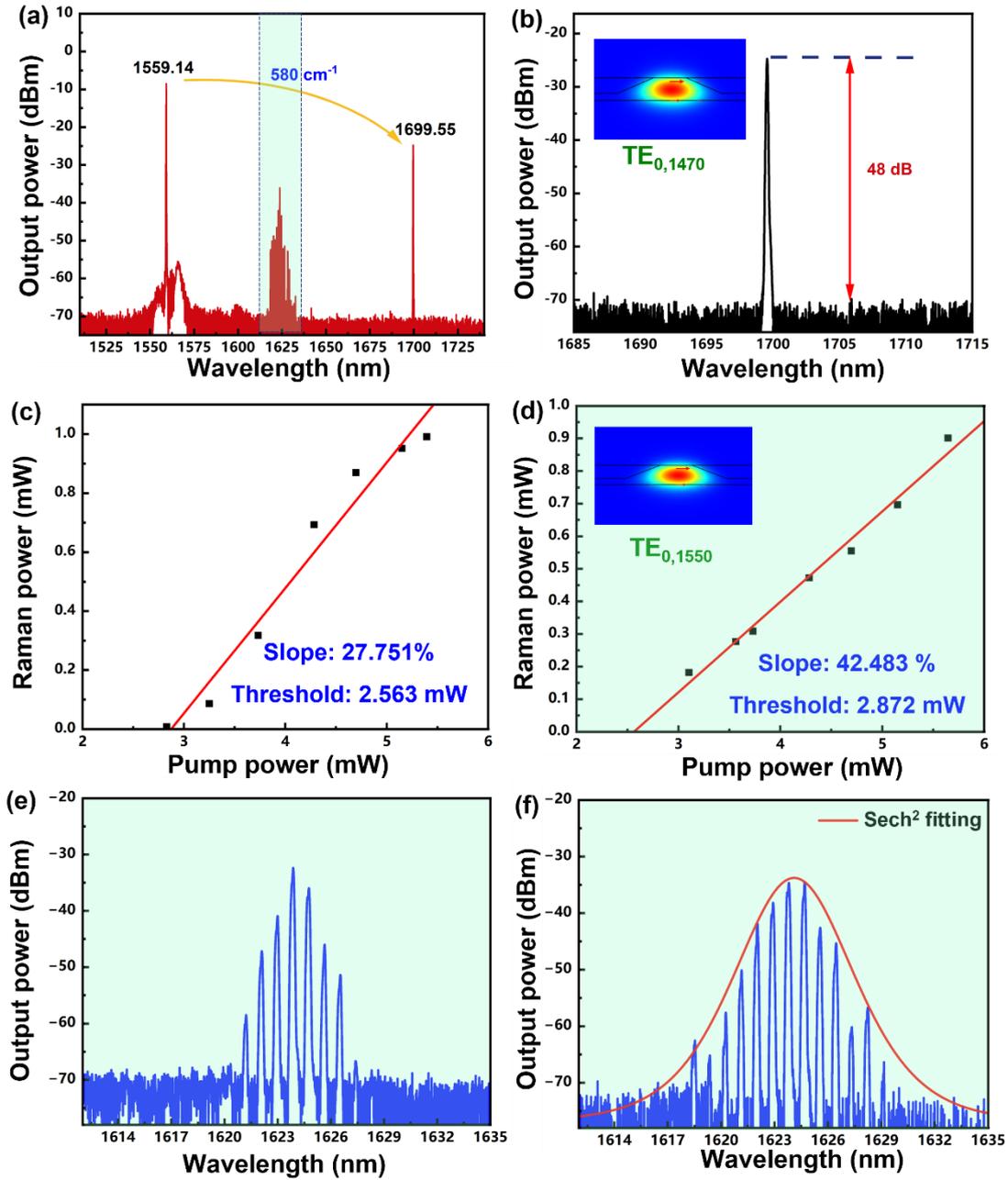

**Figure 3.** (a) Optical spectrum of the forward-propagating Stokes SRS signals around 1624 nm and 1700 nm. (b) Side-mode suppression ratio (SSR) of the SRS signal at 1699.55 nm. Inset: Simulated electric field profile of the $TE_{0,1470}$ mode at 1699.55 nm. (c) Output power of the SRS at 1699.55 nm vs. on-chip pump power. (d) Output power of the SRS at 1624.49 nm vs. on-chip pump power. Inset: Simulated electric field profile of the $TE_{0,1550}$ mode. (e) Optical spectra of the frequency comb at pump levels of (e) 3.11 mW and (f) 3.73 mW.

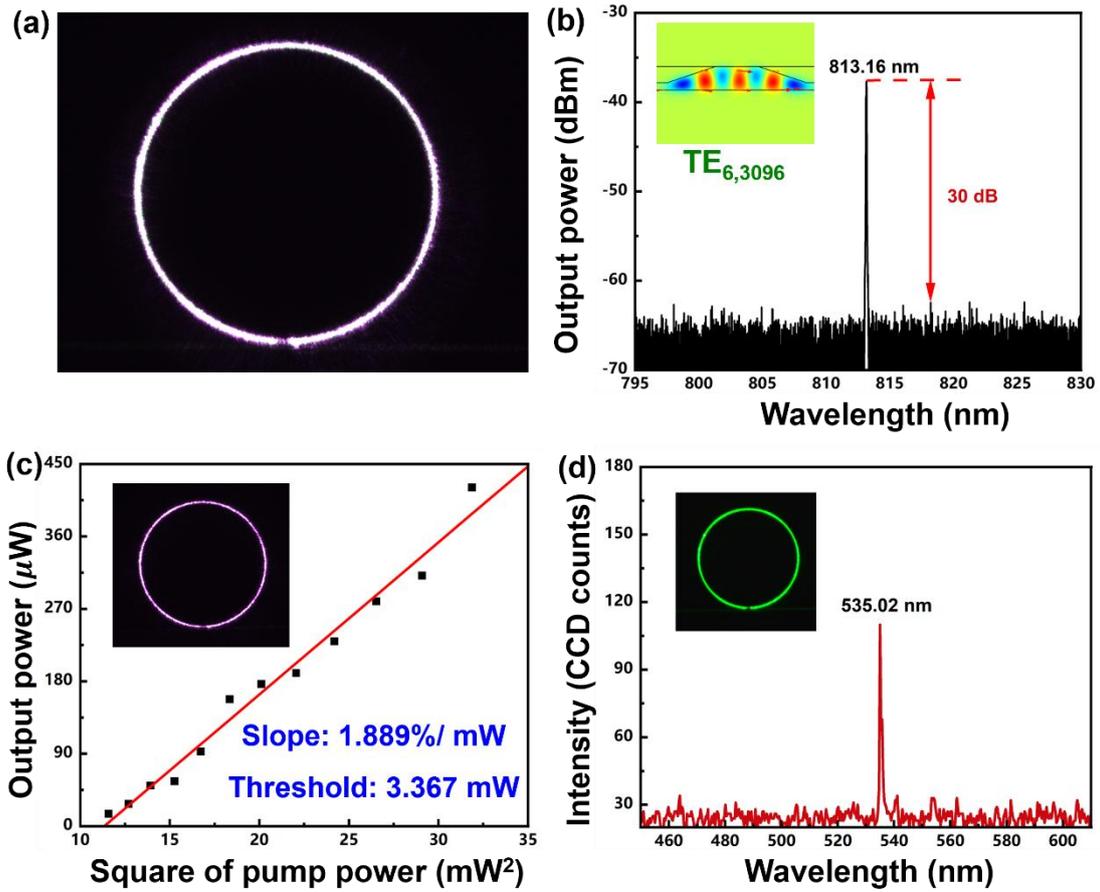

**Figure 4**. (a) Optical micrograph of nonlinear light emission from the microring captured without optical filters. (b) Optical spectrum of the near infrared Raman lasing signal (NIR-R). Inset: Simulated electric field profile of the NIR-R mode $TE_{6,3096}$. (c) Output power evolution of the NIR-R signal vs. the on-chip pump power. Inset: Optical micrograph of nonlinear light emission from the microring captured using a 950 nm short-pass and an 800 nm long-pass filter. (d) Optical spectrum of the visible Raman signal (V-R). Inset (upper): Optical micrograph of nonlinear light emission from the microring captured using a 600 nm short-pass and a 500 nm long-pass filter. Inset (lower): Optical micrograph of nonlinear light emission from the microring captured using a 450 nm short-pass filter.

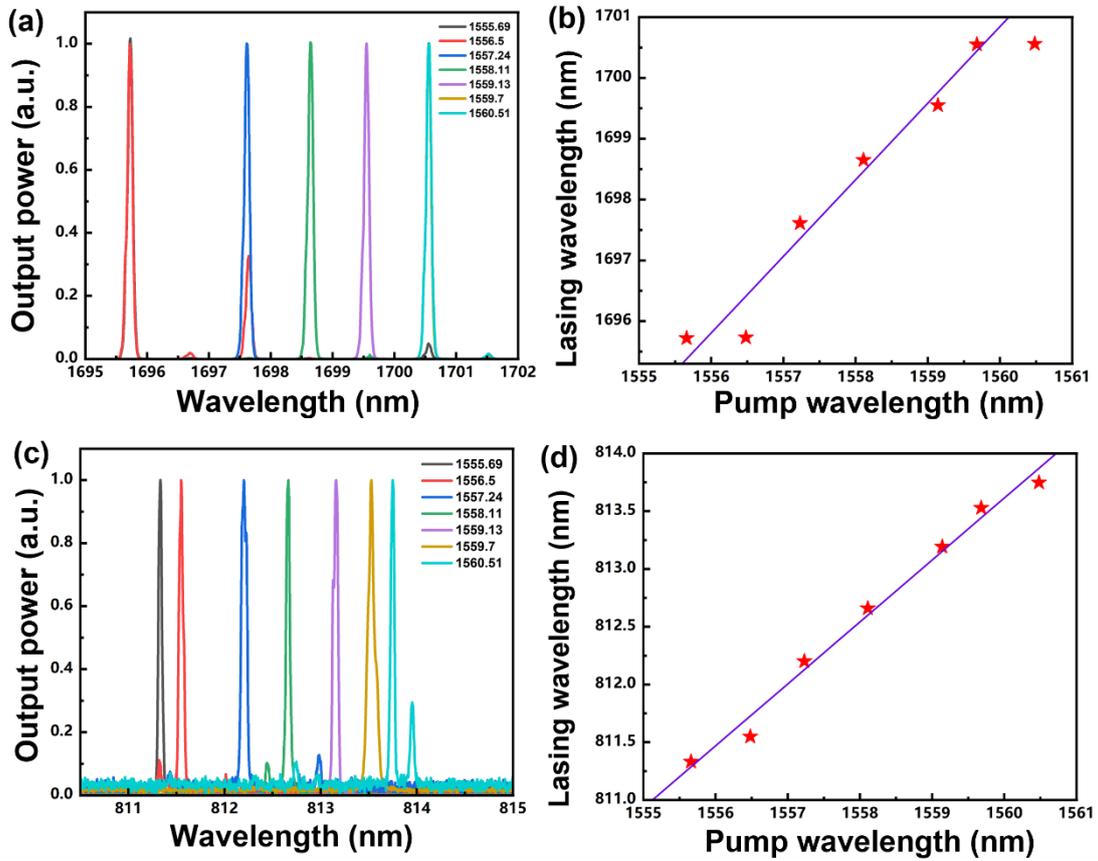

**Figure 5**. (a) Spectra of the tunable Raman laser signals. As the pump wavelength is tuned from 1555.69 nm to 1560.51 nm, the Raman laser wavelength correspondingly shifts from 1695.72 nm to 1700.56 nm. (b) Tunable Raman laser wavelength versus the pump wavelength. (c) Spectra of the tunable NIR-R signal. When the pump wavelength is tuned from 1555.66 nm to 1560.48 nm, the NIR-R laser wavelength shifts from 811.332 nm to 813.748 nm. (d) Tunable NIR-R laser wavelength versus the pump wavelength.